\documentclass[twocolumn,showpacs,preprintnumbers,amsmath,amssymb, prl]{revtex4}

\usepackage{graphicx}
\usepackage{dcolumn}
\usepackage{bm}

\begin{document}

\preprint{}

\title{Exciton spin decay modified by strong electron-hole exchange interaction}

\author{G.~V.~Astakhov$^{1,2}$}\email[E-mail: ]{astakhov@physik.uni-wuerzburg.de}
\author{A.~V.~Koudinov$^{1}$}
\author{K.~V.~Kavokin$^{1}$}
\author{I.~S.~Gagis$^{1}$}
\author{Yu.~G.~Kusrayev$^{1}$}
\author{W.~Ossau$^{2}$}
\author{L.~W.~Molenkamp$^{2}$}
\author{G.~Karczewski$^{3}$}
\author{T.~Wojtowicz$^{3}$}
\author{J.~Kossut$^{3}$}

\affiliation{$^{1}$A. F. Ioffe Physico-Technical Institute,
Russian Academy of Sciences,
194021 St. Petersburg, Russia\\
$^{2}$Physikalisches Institut der Universit\"{a}t W\"{u}rzburg,
97074 W\"{u}rzburg, Germany\\
$^{3}$Institute of Physics, Polish Academy of Sciences, 02-668
Warsaw, Poland}

\date{\today}

\begin{abstract}
We study exciton spin decay in the regime of strong electron-hole
exchange interaction. In this regime the electron spin precession
is restricted within a sector formed by the external magnetic
field and the effective exchange fields triggered by random spin
flips of the hole. Using Hanle effect measurements, we demonstrate
that this mechanism dominates our experiments in CdTe/(Cd,Mg)Te
quantum wells. The calculations provide a consistent description
of the experimental results, which is supported by independent
measurements of the parameters entering the model.
\end{abstract}


\maketitle

Spin is a fundamental invariant of an electron. In semiconductors,
the electron and hole spins can be optically oriented by a
circularly polarized light \cite{ref1}. Recently much interest has
been attracted to this field caused by possible applications
(spintronics) \cite{ref2}. In particular, the spin trapped in a
quantum dot has been proposed to serve as a qubit in the quantum
computation schemes \cite{ref3} using an all-optical protocol for
spin readout and initialization \cite{ref4}. Therefore, a detailed
understanding of the spin decay processes are of crucial
importance for the functionality of spin-based devices.

The most widespread method for studying the dynamics of the
non-equilibrium spin populations in semiconductors is optical
orientation  \cite{ref1}. Circularly polarized light excites
electrons with a predominant spin polarization along the direction of
the incident light beam.
When an external magnetic field $B$ is applied in Voigt geometry,
i.e., in the plane of the sample, the optically oriented electron
spins start precessing with Larmor frequency $\Omega = g_{e}
\mu_{B}B$ around the field direction ($g_{e}$ is the electron
g-factor). Under continuous wave (cw) excitation this results in a
decrease of the spin mean value with growing $B$, such a behavior
is known as the Hanle effect \cite{ref1}. The half width at half
maximum (HWHM) of the Hanle curve at $B=B_{1/2}$ corresponds to
the condition $\Omega T_{se} = 1$, and hence allows determination
of the electron spin lifetime $T_{se}$. For optically created
electrons it is governed by the electron spin relaxation time
$\tau_{se}$ and the lifetime time $\tau$ of the electron-hole pair
(exciton):  $T_{se}^{-1} = \tau_{se}^{-1} + \tau^{-1}$.

This classical picture will be modified drastically by
\textit{isotropic} electron-hole (eh) exchange interaction. It
manifests itself as a splitting  $\Delta_{0}$ between the
radiative $(\pm 1)$ and nonradiative $(\pm 2)$ doublets of the
neutral exciton \cite{ref5}. The electron spin now precesses in a
total field formed by the external magnetic field $B$ and the
exchange field of the hole $B_{ex} = \Delta_{0} / | g_{e} |
\mu_{B}$. Moreover, due to the hole spin relaxation, the exchange
field $B_{ex}$ changes with time introducing additional
difficulties for the theoretical description. The influence of the
eh-exchange interaction \cite{ref7,ref8} and the importance of the
hole spin relaxation rate on the electron spin decay
\cite{ref6,ref9} have been recognized in previous theoretical
works. However, the Hanle effect in the regime of the strong
electron-hole exchange coupling, when the frequencies associated
with the exchange interaction become greater than relaxation rates
of electrons and holes, has not been investigated so far. The role
of the exchange coupling is especially high in systems with
reduced dimensionality like quantum dots (QDs) or quantum wells
(QWs), where the exciton is confined in a small volume.

In this Letter we consider exciton spin decay in the regime of
strong eh-exchange interaction. We demonstrate that in this limit
the classical picture is not valid any more. While the Hanle curve
still originates from depolarization of the electron spin in an
external magnetic field, it occurs in a rather different manner
and $B_{1/2}$ now depends on the eh-exchange splitting
$\Delta_{0}$, the hole spin relaxation time $\tau_{sh}$, and the
exciton life time $\tau$. We provide strong experimental evidence
by comparing the Hanle curves of a single CdTe/(Cd,Mg)Te QW
detected at the neutral exciton ($\mathrm{X^{0}}$) and positively
charged trion ($\mathrm{T^{+}}$), where in the latter case the
eh-exchange interaction is suppressed. We find that the
characteristic depolarization field of the exciton Hanle curve
$B_{1/2}^{X}$ is an order of magnitude larger than that of the
trion Hanle curve $B_{1/2}^{T}$, indicating their different
physics. 

\begin{figure}[btp]
\includegraphics[width=.32\textwidth]{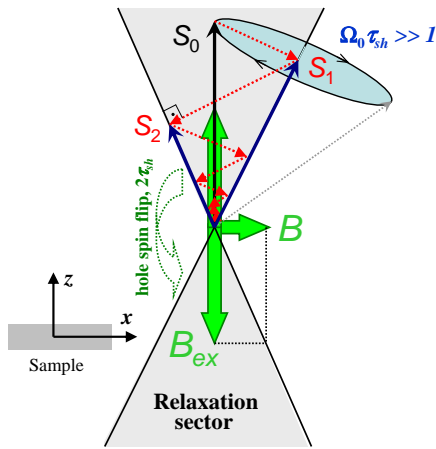}
\caption{(Color online) Chain of the electron spin decay ($S_{0}
\rightarrow S_{1} \rightarrow S_{2}...$) within the relaxation
sector shown by gray area. Such a chain represents the evolution
of the mean spin of the electron for strong exchange interaction
($\Omega_{0} \tau_{sh} \gg 1$). The relaxation sector is formed by
the two fields acting on the electron: The exchange field of the
hole $B_{ex}$ and the external magnetic field $B$. } \label{fig4}
\end{figure}

We now consider schematically the regime of strong eh-exchange
interaction  following the ideas of Ref.~\onlinecite{ref18} and
discuss the limitations of this approach (see Fig.~\ref{fig4}).
Let as an example circularly polarized light excite exciton with
the z-projection of the angular moment $J_{z} = -1$, consisting of
an electron with spin up $(+1/2)$ and a hole with spin down
$(-3/2)$. How will this system evolve in a magnetic field $B$
directed perpendicular to the quantization axis (Fig.~\ref{fig4})?

First, we assume that the in-plane hole g-factor is close to zero.
This is valid when the heavy-hole band is split off from the light
hole band as indeed occurs in low dimensional systems, and when
their mixing may be neglected. We also ignore the anisotropic
exchange interaction \cite{ref5}. Under this assumption the hole
spin is always directed parallel to the z-axis (perpendicular the
sample plane), and the spin dynamics the hole is capable of are
random spin flips with a mean time between the successive flips of
$2 \tau_{sh}$. In other words, this applies a hole spin relaxation
with a characteristic time $\tau_{sh}$.

The electron, possessing a nonzero g-factor, experiences the
action of the external field $B$ and of the exchange field of the
hole $B_{ex} = \Delta_{0} / | g_{e} | \mu_{B}$, which in our
example is initially directed upwards (Fig.~\ref{fig4}). The
electron spin precesses in the resulting total field, the
condition of strong eh-exchange implying that the Larmor frequency
$\Omega_{0} = \Delta_{0} / \hbar$ is so high that the electron
spin makes at least one turn around the total field before the
precession is interrupted by a flip of the hole spin. Formally,
this condition reads $\Omega_{0} \tau_{sh} > 1$. If this is
fulfilled, only the projection $S_{1}$ of the initial mean
electron spin $S_{0}$ on the direction of the total field is
conserved until the moment of the hole spin flip, while the
transverse components of the electron spin become random
(Fig.~\ref{fig4}).

The above process is only a first link in the chain of the spin
evolution of the exciton. The flip of the hole spin results in the
inversion of the direction of the exchange field; the direction of
the total field will consequently change and the electron spin
precession will proceed around the new direction (see
Fig.~\ref{fig4}). Analogously to the previous consideration, the
precession will lead to a transformation of the mean spin $S_{1}$
into $S_{2}$. After that, the new flip of the hole spin will
return the total field to the initial direction. The next step
will result in the decay of the new portion of the electron spin,
etc. (Note that we implicitly assume a long electron spin
relaxation time $\tau_{se} > \tau_{sh}$.) The cascade process of
the spin depolarization of the electron will last until it is
interrupted by excitonic recombination with a characteristic time
$\tau$.

As a result of the above deliberation, we can introduce a
relaxation sector shown by the gray area in Fig.~\ref{fig4}, which
lies in the plane formed by $B$ and $B_{ex}$ fields. The
relaxation sector restricts the directions of the electron spin
and determines its decay path. This picture is thus valid for zero
in-plane hole g-factor, and provided $1/\Omega_{0} < \tau_{sh} <
\tau_{se}, \tau$. As we show in the following, these conditions
can indeed be realized experimentally.

The sample under study was grown by molecular beam epitaxy on a
(001) GaAs substrate. The structure consists of a single
(38-\AA-wide) $\mathrm{CdTe/Cd_{0.7}Mg_{0.3}Te}$ QW and is
nominally undoped. The photoluminescence (PL) was excited by a dye
laser with tunable photon energy and detected with a 1-m
spectrometer and a photomultiplier tube. In order to detect the
polarization of the emission we used a standard scheme with a
piezo-elastic modulator and a two-channel photon counter. Magnetic
field was applied either perpendicular to the sample plane
(Faraday geometry) or in the sample plane (Voigt geometry). All
experiments were carried out at a temperature 1.6~K.

\begin{figure}[tbp]
\includegraphics[width=.41\textwidth]{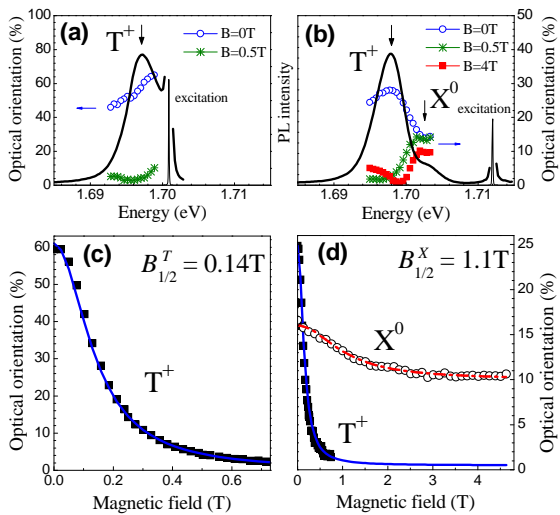}
\caption{(Color online) Spectra of PL and optical orientation for
quasi-resonant excitation at the positively charged trion (a) and
non-resonant excitation above the neutral exciton (b). (c) The
Hanle effect measured for the trion ($\mathrm{T^{+}}$) at the same
excitation conditions as in panel (a). Solid line is the fit using
Eq.~(\ref{Eq1}) with characteristic field $B_{1/2}^{T}= 0.14$~T.
(d) The Hanle effect measured for the exciton ($\mathrm{X^{0}}$)
at the same excitation conditions as in panel (b). Dashed line is
the fit using Eqs.~(\ref{Eq6}) and (\ref{Eq7}) with characteristic
field $B_{1/2}^{X}= 1.1$~T. Magnetic fields are applied in Voigt
geometry.} \label{fig1}
\end{figure}

\begin{figure}[tbp]
\includegraphics[width=.41\textwidth]{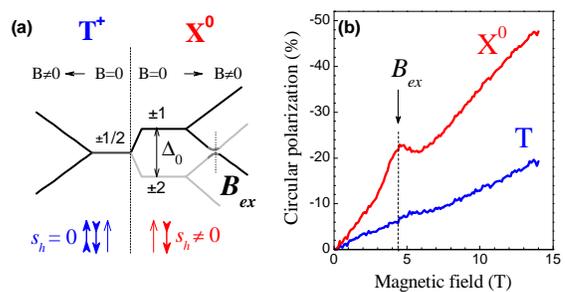}
\caption{(Color online) (a) Energy diagrams of the trion
($\mathrm{T^{+}}$) and exciton ($\mathrm{X^{0}}$) spin levels in a
magnetic field applied in Faraday geometry. In the studied samples
the isotropic exchange splitting is $\Delta_{0} = \hbar \Omega_{0}
\approx 0.33$~meV. For an external magnetic field $B = B_{ex}$ the
radiative (black) and nonradiative (gray) branches exhibit
anti-crossing. (b) Circular polarization of the exciton
($\mathrm{X^{0}}$) and the trion ($\mathrm{T}$) emission induced
by the magnetic field at a non-resonant excitation (above the QW
barrier). At  $B_{ex}
\approx 4.5$~T a distinct anomaly is observed. 
} \label{fig2}
\end{figure}

A typical PL spectrum is shown in Fig.~\ref{fig1}(b). Generally,
the spectrum consists of two bands. The high-energy peak is
attributed to the neutral exciton $\mathrm{X^{0}}$, and the
low-energy one is ascribed to a charged exciton, or trion,
$\mathrm{T}$ \cite{ref11}. Our analysis shows that we are dealing
with the positively charged trion $\mathrm{T^{+}}$ \cite{ref12}.
It consists of one electron and two holes. The holes form a spin
singlet with zero total spin $s_{h} = 0$ as shown in
Fig.~\ref{fig2}(a), so the optical orientation signal from
$\mathrm{T^{+}}$ always reflects the spin polarization of the
unpaired electron $s_{e}$. Moreover, for the same reason (i.e.,
$s_{h} = 0$) the exchange field is zero for a $\mathrm{T^{+}}$
trion, $B_{ex} = 0$, which allows us to study the electron spin
relaxation free of exchange interaction.

Figure~\ref{fig1}(a) shows a typical spectrum of the optical
orientation under quasi-resonant excitation at the trion line.
While quite high in absence of external field, the optical
orientation signal approaches zero already at a relatively low
magnetic field of $B=0.5$~T. The Hanle curve detected at the PL
maximum is given in Fig.~\ref{fig1}(c) by squares. It has a
Lorentzian shape and is perfectly described by the classical
formula derived for the free spin precession of an electron
\cite{ref1}
\begin{equation}
P_{T^{+}} = 2 s_e = \frac{T_{se}}{\tau_r} \frac{2
s_{e0}}{1+(B/B_{1/2}^{T})^2} \,\,, \label{Eq1}
\end{equation}
with $s_e$ being its mean spin value and
\begin{equation}
B_{1/2}^{T} = \frac{\hbar}{|g_e| \mu_B T_{se}} \,\,\, \mathrm{and}
\,\,\, \frac{1}{T_{se}} =  \frac{1}{\tau_{r}} +
\frac{1}{\tau_{se}} \,. \label{Eq2}
\end{equation}
Here $s_{e0}$ is the initial value of the electron spin as created
at the $\mathrm{T^{+}}$ state by optical orientation. The best fit
to the data in Fig.~\ref{fig1}(c) is achieved with $2 s_{e0}
T_{se} / \tau_{r} = 0.61$ and  $B_{1/2}^{T}= 0.14$~T (solid line).
Taking the electron g-factor $g_{e}= -1.3$ known from the
spin-flip Raman scattering experiments \cite{ref13} and the
obtained value of $B_{1/2}^{T}$, we deduce for the electron spin
lifetime $T_{se}=61$~ps. Assuming that the conditions of
quasi-resonant excitation are close to the ideal case ($s_{e0}
\approx 1/2$), one can estimate a recombination time of
$\tau_{r}=100$~ps. This value is in agreement with typical decay
times of the trion PL obtained from direct time-resolved
experiments on CdTe-based QWs \cite{ref14}. In the case of
nonresonant excitation [Fig.~\ref{fig1}(b)] the HWHM of the trion
Hanle curve [Fig.~\ref{fig1}(d)] remains unchanged, again yields
$B_{1/2}^{T}= 0.14$~T [Fig.~\ref{fig1}(d)]. Reduction of the
initial electron spin ($s_{e0} < 1/2$) is most likely caused by
relaxation processes to the trion ground state.

In contrast to the positively charged trion, the neutral exciton
$\mathrm{X^{0}}$ displays a rather different behavior [see
Fig.~\ref{fig1}(d)]. First, while the magnetic field is increased,
the polarization does not disappear but saturates at a nonzero
level of $\sim 0.1$. Second, while there exists a field-dependent
part of polarization which is well fitted by a Lorentzian curve,
analogously to the case of the trion, the characteristic field of
this depolarization equals $B_{1/2}^{X} = 1.1$~T, much larger than
that of the $\mathrm{T^{+}}$.

Give the above discussion, the natural explanation of this
observation is that the wide depolarization curve of the exciton
originates from the exchange interaction between the electron and
the hole (as in Fig.~\ref{fig4}). In order to examine this we
measured the exchange splitting $\Delta_0$ independently.
Following scheme shown in Fig.~\ref{fig2}(a), a magnetic field $B$
applied in the Faraday configuration may lead at some field values
to an anti-crossing of the bright and dark exciton states
\cite{ref15}. The first anti-crossing occurs when $|g_{e}| \mu_{B}
B = \Delta_{0}$, and hence this magnetic field $B = B_{ex}$ may be
interpreted without any additional assumption as the exchange
field of the hole acting on the electron . Experimentally, the
anti-crossing manifests itself as an anomaly in the polarization
of the PL \cite{ref16,ref17}. Indeed, we observed such an anomaly
(at $B_{ex} \approx 4.5$~T) when detect circular polarization
degree of the exciton in emission [Fig.~\ref{fig2}(b)]. Thus we
have the following relationship between characteristic magnetic
fields $B_{1/2}^{T} \ll B_{1/2}^{X} \ll B_{ex}$.

We now demonstrate that the experimental behavior is in good
agreement with an analytical solution for the model presented in
Fig.~\ref{fig4}. Assuming $\tau_{sh} \ll \tau$ the electron
depolarization essentially occurs for weak external fields $B \ll
B_{ex}$. So the relaxation sector shown in gray in Fig.~\ref{fig4}
is narrow, and the decrease of the electron spin density $S_{e}$
per single flip of the hole spin approximately equals $2 S_{e} (B
/ B_{ex})^2$. The master equation for $S_{e}$ then takes a simple
form
\begin{equation}
\dot{S_e} = s_{e0} G - \frac{S_e}{\tau}  - \frac{S_e}{\tau_{sh}}
\left( \frac{B}{B_{ex}} \right)^2 \,; \label{Eq3}
\end{equation}
here the first term describes the spin creation with the
excitation rate $G$ (with $s_{e0} \leq 1/2$  phenomenologically
accounting for possible losses of the electron spin polarization
in the higher-lying state where the exciton is actually created),
the second term represents the losses by excitonic recombination,
the third term describes the decrease of spin via the chain
process of Fig.~\ref{fig4} [$(2 \tau_{sh})^{-1}$ has the sense of
number of hole spin flips per unit of time]. Setting $\dot{S_e}$
to zero, one can find the steady-state value of $S_{e}$  as a
function of $B$
\begin{equation}
S_e = s_{e0} \frac{G \tau}{1 + \frac{\tau}{\tau_{sh}} \left(
\frac{B}{B_{ex}} \right)^2} \,. \label{Eq4}
\end{equation}
When $\tau_{sh} \ll \tau$, the correlation between the electron
and hole spins is weak, so the intensities of the components of
the luminescence with opposite circular polarizations ( $I_{+}$
and $I_{-}$) depend only on the mean spin values  $s_{e} = S_{e} /
G \tau$ and $s_{h} = - \frac{3}{2} \tau_{sh} / \tau$:
\begin{equation}
I_{\pm} \propto \frac{1}{3} \left( \frac{1}{2} \pm s_e \right)
\left( \frac{3}{2} \mp s_h \right)  \,. \label{Eq5}
\end{equation}
Using once more that $\tau_{sh} / \tau$  is small,  we finally
obtain the luminescence polarization:
\begin{equation}
P_{X^{0}} \approx  \frac{\tau_{sh}}{\tau} + \frac{2 s_{e0}}{1 +
\frac{\tau}{\tau_{sh}} \left( \frac{B}{B_{ex}} \right)^2} \,.
\label{Eq6}
\end{equation}
The first (field-independent) contribution originates from the
mean spin polarization of holes. Its value is controlled by the
ratio of the hole spin relaxation time and the exciton lifetime.
The inverse of this ratio also enters the second term describing
the chain spin depolarization of electrons.

We now compare the result of the calculation with the experimental
data. The ratio of the hole spin relaxation time and the exciton
lifetime can be estimated from the high-field value in
Fig.~\ref{fig1}(c), $\tau_{sh} / \tau \approx 0.1$ . This estimate
looks reasonable, as for CdTe-based QWs the typical reported
values of the hole spin relaxation time are on the order of tens
of picoseconds \cite{ref19}. Together with the value of the
exchange field $B_{ex} \approx 4.5$~T taken from the experimental
data in Fig.~\ref{fig2}(b), this gives a Lorentzian depolarization
curve with a HWHM of 1.4~T, in reasonable agreement with the
experimentally measured value $B_{1/2}^{X}=1.1$~T. It turns out
that in the strong exchange interaction regime, multiple flips of
the hole spin result in a narrowing of the depolarization curve by
a factor $\sqrt{\tau / \tau_{sh}} \approx 3.2$ as compared to the
field of the exchange interaction:
\begin{equation}
B_{1/2}^{X} = B_{ex} \sqrt{\tau_{sh} / \tau} \,. \label{Eq7}
\end{equation}
Equation~(\ref{Eq7}) summarizes our key result. Note that if the
exciton lifetime $\tau$ is controlled by the radiative
recombination time $\tau_r$, as for the trion ( $\tau \approx
\tau_r = 100$~ps), then $\tau_{sh} \approx 10$~ps and the
condition of strong exchange interaction is reasonably satisfied,
$\Omega_0 \tau_{sh} = 5$.

In summary, by studying the depolarization of the PL from of the
positive trion and from the exciton in a CdTe QW, we were able to
investigate the exciton spin decay subject to strong exchange
interaction. The spin of an electron interacting with a hole
precesses in tilted magnetic fields, trigged by the random spin
flips of the hole. This process results in a depolarization
pattern with the characteristic field $B_{1/2}^{X} = \Delta_0
\tau^{1/2} / |g_e| \mu_B \tau_{sh}^{1/2}$, which was
experimentally observed and verified by separate determination of
$\Delta_0 / |g_e| \mu_B$ and $\tau^{1/2} / \tau_{sh}^{1/2}$ . The
reported mechanism is of interest for nanostructures with sizable
isotropic electron-hole exchange interaction (e.g., $\Delta_0
\approx 0.33$~meV in this work).

The study was supported by INTAS (03-51-5266) as well as by Russian Foundation for Basic Research and SFB 410.



\begin{references}
\bibitem{ref12} Identification whether the trion is positively or negatively charged
is based on experimental value of the characteristic magnetic
field $B_{1/2}^{T} = 0.14$~T [Fig.\ref{fig1}(c)]. The Hanle effect
detected on the trion line at a quasi-resonant excitation must
arise from the precession of spin of the unpaired particle. That
particle can not be a hole, as the hole possesses too small
in-plane g-factor $g_{h\bot} < 0.04$ \cite{ref13} resulting in the
spin lifetime $T_{s} > 2$~ns. The latter value is clearly
meaningless in view of typical recombination times $\sim 100$~ps
reported for similar QWs in many papers. So the particle
responsible for the low-energy PL line must be positively charged
trion $\mathrm{T^{+}}$ which includes unpaired electron.
\end{references}
\end{document}